\pdfoutput=1

\documentclass[conference]{IEEEtran}
\ifCLASSINFOpdf
\else
\fi
\usepackage{amsmath, amssymb}
\usepackage{graphicx}
\usepackage{caption}
\usepackage{subcaption}
\usepackage{algorithm,float}
\usepackage{algpseudocode}
\begin{document}

\title{Investigation of Holographic Beamforming via Dynamic Metasurface Antennas in QoS Guaranteed Power Efficient Networks}

\author{
    \IEEEauthorblockN{Askin Altinoklu, and Leila Musavian}
    \IEEEauthorblockA{
        CSEE, University of Essex, UK\\
        Emails: \{askin.altinoklu, leila.musavian\}@essex.ac.uk}
}


%Telephone: (800) 555--1212, Fax: (888) 555--1212}
\bibliographystyle{IEEEtran}


\maketitle

\begin{abstract}
This work focuses on designing a power-efficient network for Dynamic Metasurface Antennas (DMA)-aided multi-user multiple-input single-output (MISO) antenna systems. Power efficiency is achieved through holographic beamforming in a DMA-aided network, minimizing total transmission power while ensuring a guaranteed signal-to-noise-and-interference ratio (SINR) for multiple users in downlink. Unlike conventional MISO systems, which have well-explored beamforming solutions, DMA require specialized methods due to their unique physical constraints and wave-domain precoding capabilities. To achieve this, optimization algorithms relying on alternating optimization and semi-definite programming, are developed, including spherical-wave channel modelling of near-field communication. In this setup, the beamforming performance of DMA-aided precoding is analyzed in comparison to its optimal limits and traditional fully digital (FD) architectures, considering the effects of the Lorentzian constraints of metasurfaces and the degree of freedom (DoF) limitations due to a reduced number of RF chains. We demonstrate that the performance gap caused by DoF constraints becomes more significant as the number of users increases, highlighting the trade-offs of DMA in high-density wireless networks.
\end{abstract}


% \fi
\IEEEpeerreviewmaketitle

\section{Introduction}
%As the transition from 5G to 6G for mobile communication and internet of things networks is now rapidly progressing, key performance indicators (KPIs) and requirements for the communication standards of 6G networks became more apparent. It is well anticipated that the demand for data rate related KPIs such as the user experienced data rate and peak data rate will be among major concerns of 6G, as it was for 5G \cite{6g_kpi}. On the other hand, limitations on the overall power consumption of the future communication networks will be a significant concern under the paradigm shift toward green communications \cite{6g_green}.
%Considering that the number of devices involved in communication networks with various Quality of Services (QoS) constraints will be increased dramatically, and the typical data rates of future applications will be much higher than those of today, usage of distributed antenna apertures together with multiple-input multiple output (MIMO) and/or multiple-input single output (MISO) transmission will be inevitable for the future communication networks. 
As the transition from 5G to 6G for mobile communication and Internet of Things networks is rapidly progressing, key performance indicators (KPIs) and requirements for 6G communication standards have become more apparent. The demand for data rate-related KPIs, such as user-experienced and peak data rates, will remain a major concern for 6G, as it was also for 5G \cite{6g_kpi}. Considering the dramatic increase in devices with various Quality of Service (QoS) constraints and the significantly higher data rates of future applications, the use of distributed antenna apertures with multiple-input multiple-output (MIMO) and/or multiple-input single-output (MISO) transmission will be inevitable for future communication networks. Meanwhile, with the effect of newly emerging technologies,  distributed antenna arrays used within traditional MIMO/MISO systems are at the edge of transition from static predesigned intrinsic elements to dynamically reconfigurable array elements such as metasurfaces \cite{metasurface_6g}. The dynamic reconfigurability of metasurfaces, achieved through their physical properties such as permittivity and permeability, provides flexibility on the control and manipulation of electromagnetic (EM) waves that interact with them. Leveraging this characteristic, several distinct types of metasurface arrays have been developed in various applications of MIMO and MISO systems, whether as transmitter, receiver or within the wireless channel. Particularly, Dynamic Metasurface Antennas (DMA) are shown to be an effective solution for downlink and uplink  beamforming problems in MIMO and MISO systems \cite{metasurface_6g, DMA1}. Taking the leverage of wave-domain precoding, DMA require less number of RF chains than the conventional MIMO and/or MISO architectures, such as an FD architecture, and do not need any additional analog circuitry such as phase shifters as in hybrid arcitectures \cite{DMA_BF1}. These features make DMA a remarkable candidate for the future communication networks. Recent studies based on DMA have presented the effectiveness of beamforming in multiuser MISO downlink systems in various applications \cite{DMA1, DMA_BF1,Azarbahram_2024}. 

%On the other hand, the physical constraints on DMA weights and wave-domain precoding of multiple DMA elements through reduced number of RF suppliers can limit the degrees of freedom (DoF) in beamforming optimizations compared to conventional fully digital (FD) architectures. Although increasing the number of DMA elements, with its ability to decrease element spacing below $\lambda$/2 limit of conventional arrays, can alleviate these limitations, the impact on beamforming is not well-understood yet. In \cite{DMA}, the performance gap of DMA-aided precoding to the fundamental theoretical limits were examined in MIMO networks for parameters such as, signal-to-noise ratio (SNR) and number of RF chains. The comparisons of the beamforming performance of DMA to the FD have been presented for the multiuser MISO networks in various applications such as achievable sum-rate maximization in \cite{DMA1} and transmitted power minimization \cite{10339299}. However, the effect of DoF limitations on beamforming, particularly its dependency on the number of users, has not been systematically investigated. This work addresses this gap by analyzing DoF effects in DMA-aided power-efficient networks under signal-to-noise-and-interference ratio (SINR) constraints. 
On the other hand, the physical constraints on DMA weights, i.e., Lorentzian-constrained polarizabily, and wave-domain precoding via reduced RF chains can limit the degrees of freedom (DoF) in beamforming optimization compared to fully digital (FD) architectures. In \cite{DMA1}, the performance gap of DMA-aided precoding compared to the fundamental theoretical limits were examined in MIMO networks for parameters such as, signal-to-noise ratio (SNR) and number of RF chains. The comparisons of the beamforming performance of DMA and FD have been conducted for multi-user MISO networks focusing on sum-rate maximization \cite{DMA1} and transmitted power minimization \cite{Azarbahram_2024}. However, the effect of DoF limitations on beamforming, particularly its dependency on the number of users, has not been systematically investigated. This work addresses this gap by analyzing DoF effects in DMA-aided power-efficient networks under signal-to-noise-and-interference ratio (SINR) constraints. Our main contributions are as follows: i) we propose a novel holographic beamforming method based on semi-definite relaxation for minimizing transmitted power in SINR guranteed multiuser MISO networks, and demonstrate its robustness across various scenarios, ii) we perform comprehensive Monte-Carlo simulations with randomly located users for different scenarios, to highlight the impact of Lorentzian-constrained polarizability and reduced DoF on the optimal beamforming and the transmitted power in comparison to the unconstrained generic optimization problems,  iii) we analyze these effects across varying parameters, such as user position, number of users, and number of antenna, in comparison to the results of FD architecture, and iv) we demonstrate the dependency of the performance gap on the number of users, providing insights into dense user setups for future communication networks.

%The results reveal that the performance gap caused by DoF constraints remains constant for a fixed SINR with a given number of users but increases as the number of users increase.
%\subsection*{\hspace{0.5cm}Notations:}
%A matrix is denoted by the boldface capital \(\mathbf{W}\) with \(\text{rank}(\mathbf{W})\), \(\text{Tr}(\mathbf{W})\), \(\mathbf{W}^T\), \(\mathbf{W}^H\), and \(\text{Vec}(\mathbf{W})\) representing its rank, trace, transpose, Hermitian conjugate, and vectorization, respectively. A vector is denoted by the boldface lower-case \(\mathbf{w}\), with \(\|\mathbf{w}\|\), \(\mathbf{w}^T\), and \(\mathbf{w}^H\) indicating its Euclidean norm, transpose, and Hermitian conjugate, respectively. The unit vector in the direction of \(\mathbf{w}\) is denoted by \(\hat{\mathbf{w}}\), while the scalar is represented by \(w\), and \(|w|\) is its absolute value. The Kronecker product is indicated by \(\otimes\).

%\textbf{Notations:} A matrix is denoted by \(\mathbf{W}\), with \(\text{rank}(\mathbf{W})\), \(\text{Tr}(\mathbf{W})\), \(\mathbf{W}^T\), \(\mathbf{W}^H\), and \(\text{Vec}(\mathbf{W})\) representing its rank, trace, transpose, Hermitian conjugate, and vectorization, respectively. A vector is denoted by \(\mathbf{w}\), with \(\|\mathbf{w}\|\), \(\mathbf{w}^T\), and \(\mathbf{w}^H\) indicating its Euclidean norm, transpose, and Hermitian conjugate. The unit vector in the direction of \(\mathbf{w}\) is \(\hat{\mathbf{w}}\), while a scalar is \(w\), and \(|w|\) is its absolute value. The Kronecker product is denoted by \(\otimes\).
\textbf{Notations:} Matrices are denoted by \(\mathbf{W}\), with  \(\text{Tr}(\mathbf{W})\), \(\mathbf{W}^T\), \(\mathbf{W}^H\), and \(\text{Vec}(\mathbf{W})\) representing, trace, transpose, Hermitian conjugate, and vectorization. Vectors are \(\mathbf{w}\), with \(\|\mathbf{w}\|\), \(\mathbf{w}^T\), and \(\mathbf{w}^H\) indicating Euclidean norm, transpose, and Hermitian conjugate. Scalars are \(w\), with \(|w|\) as the absolute value. The Kronecker product is denoted by \(\otimes\).

\section{SYSTEM MODEL \& PROBLEM FORMULATION}
\begin{figure}[!t]
    \centering
    \begin{subfigure}[b]{0.40\textwidth}
        \centering
        \includegraphics[trim=10 30 0 30, clip, width=\textwidth]{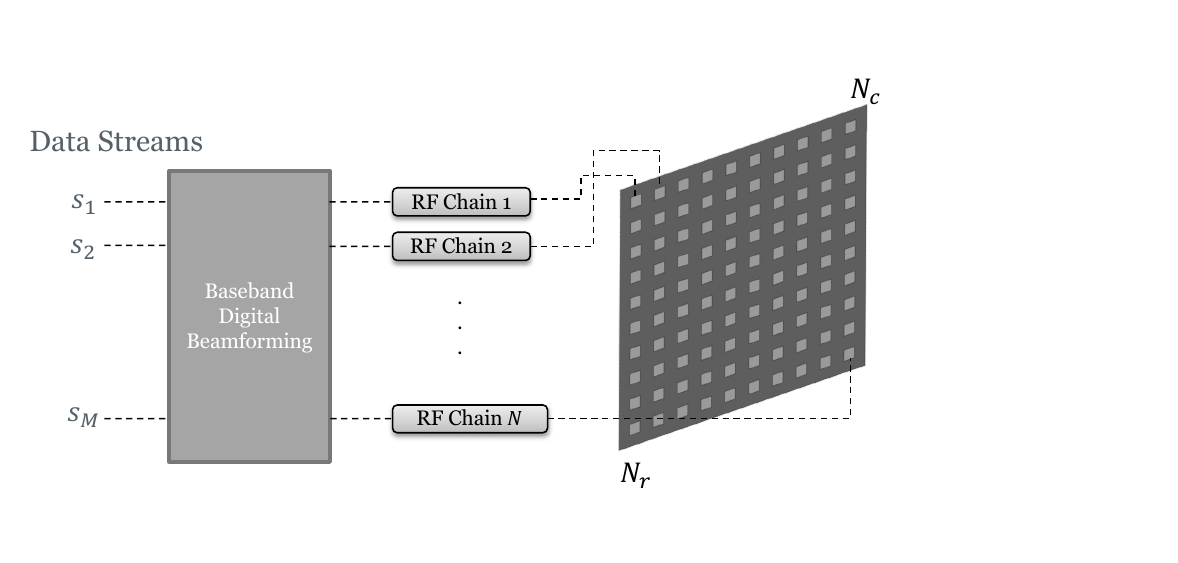}
        \caption{}
        \label{fig:sub1}
    \end{subfigure}
    \hfill
    \begin{subfigure}[b]{0.40\textwidth}
        \centering
        \includegraphics[trim=10 30 0 30, clip, width=\textwidth]{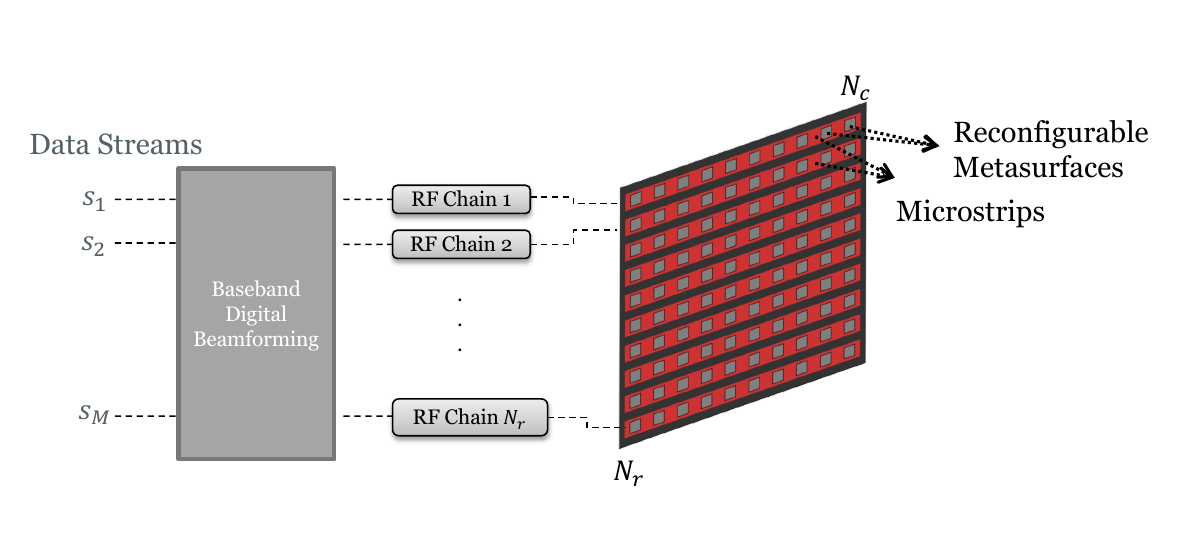}
        \caption{}
        \label{fig:sub2}
    \end{subfigure}
    \caption{System configurations for different architectures: (a) FD, (b) DMA.}
    \label{fig:main}
\end{figure}

We consider a multi-user downlink MISO\footnote{We use the term MISO since receiver users have a single antenna. However, DMA-aided MISO networks are often termed MIMO, as DMA act as HMIMO surfaces with MIMO-like two-dimensional precoding although a single antenna receiver is considered \cite{DMA_BF1}.} system, where the base station (BS) is equipped with either FD-aided (Fig. \ref{fig:sub1}) or DMA-aided (Fig. \ref{fig:sub2}) beamforming architectures. In this system, the BS serves \(K\) users, each equipped with a single antenna and requesting a certain level of SINR, in a generic Line-of-Sight (LoS) channel model that covers both far-field and near-field users. Moreover, perfect knowledge of the channel is assumed. 
\subsection{Channel and Path-Loss Modeling}
%In order to obtain channel modeling, we consider modeling with a uniform planar array (UPA) consisting of \(N_\text{r}\) and \(N_\text{c}\) subwavelength elements in the vertical and horizontal directions, respectively. This yields a total of \(N = N_\text{r}N_\text{c}\) radiating subwavelength elements. The radiation pattern \(G_\text{e}(\psi)\) of individual elements above a conducting plate is well-approximated with (\cite{gain_element}):
We consider a uniform planar array (UPA) consisting of \(N_\text{r}\) and \(N_\text{c}\) sub-wavelength elements in the vertical and horizontal directions, yielding a total of \(N = N_\text{r}N_\text{c}\) elements. The radiation pattern of the elements is well-approximated with:
\begin{equation}
\label{eq:1}
G_\text{e}(\psi) = 
\begin{cases} 
2(g+1) \cos^{g}(\psi), & 0 \leq \psi \leq \pi/2 \\
0, & \pi/2 < \psi \leq \pi 
\end{cases}
\end{equation}
where $\psi$ is the angle measured from the boresight of the array, \(g\) is the measure of antenna gain \cite{gain_element}. In our problem, we consider subwavelength elements as dipoles, i.e., \(g=2\). 
The path loss for the user positioned at $\mathbf{r}_k$ due to the ($i$,$l$)-th element of the UPA (\(\mathbf{r}_{i,l}\)) can be expressed with:
\begin{equation}
\label{eq:2}
\gamma_{k}(i, l) = \sqrt{G_\text{e}(\psi)} \frac{\lambda}{4\pi \|\mathbf{r}_k - \mathbf{r}_{i, l}\|} e^{-j \beta_0 \|\mathbf{r}_k - \mathbf{r}_{i, l}\|},
\end{equation}
where \(\lambda\), and \(\beta_0\) are the free-space wavelength and wavenumber, respectively. The entries \(\gamma_{k}(i, l)\) are the elements of the channel vector \(\boldsymbol{\gamma}_k \in \mathbb{C}^{N \times 1}\), i.e., \(\boldsymbol{\gamma}_k \triangleq \begin{bmatrix}
\gamma_{k}(1, 1), \gamma_{k}(1, 2), \ldots, \gamma_{k}(N_\text{r}, N_\text{c})
\end{bmatrix}^H\).

\subsection{DMA Architecture}

The DMA-based architecture comprises of \(N_\text{r}\) microstrips each containing \(N_\text{c}\) metasurface-based intrinsic elements, yielding \(N \triangleq N_\text{c} N_\text{r}\) of total elements. The metasurface elements are excited by guided (reference) waves propagating through each associated microstrip. Each microstrip is connected to the digital beamformer through a single RF chain and the complex amplitudes of the reference wave for each microstrip are controlled by digital beamforming vectors \(\mathbf{w}_m \in \mathbb{C}^{N_\text{r} \times 1}\) for all \(m \in M\), where \(M = \min(K, N_\text{r})\). The radiated pattern of individual elements is further adjusted with their corresponding dynamically configurable weights (\(\mathbf{Q} \in \mathbb{C}^{N \times N_\text{c}}\)). Hence, the transmitted signal for DMA can be defined as
\begin{equation}
\mathbf{x} = \sum_{m=1}^{M} \mathbf{x}_m = \sum_{m=1}^{M} \mathbf{H} \mathbf{Q} \mathbf{w}_m s_m.
\label{eq:6}
\end{equation}
In (\ref{eq:6}), \(\mathbf{H}\) is a diagonal matrix with elements \(\mathbf{H}_{(i-1)N_\text{c} + l, (i-1)N_\text{c} + l} = e^{-d_{i,l} (\alpha_i + j \beta_i)}\), where \(d_{i,l}\) is the position of the \(l\)-th element along the \(i\)-th microstrip, and \(\alpha_i, \beta_i\) are its attenuation and propagation constants \cite{metasurface_6g}. Matrix \(\mathbf{Q}\) includes the inter-connectivity of the individual elements to the excitation ports, with their frequency-dependent response to the external excitation. This response can be well-defined by the Lorentzian resonance response, i.e., 
\begin{equation}
\label{eq:7p}
\mathbb{Q} = \left\{ q = \frac{j + e^{j\Phi}}{2} : \Phi \in [0, 2\pi] \right\}.
\end{equation} 
Matrix \(\mathbf{Q}\) can therefore be defined in a block-diagonal form, as described in \cite{DMA_BF1} with
\begin{equation}
\label{eq:7}
\mathbf{Q}_{(i-1)N_\text{c} + l, n} =
\begin{cases}
q_{i,l} \in \mathbb{Q}  & \text{if } i = n \\
0 & \text{if } i \ne n.
\end{cases}
\end{equation}
Finally, by inserting the antenna aperture distributions from (\ref{eq:6}) into the channel model given in (\ref{eq:2}), the received signal at User \(k\) can be formulated as
\begin{equation}
\label{eq:8}
{y}_{k} = \boldsymbol{\gamma}_k^H \sum_{m=1}^{M} \mathbf{H} \mathbf{Q} \mathbf{w}_m s_m + n_k.
\end{equation}
\subsection{General Problem Formulation}
%Following the derivations of transmitted and received signals \(\mathbf{x}_k\), \(\mathbf{y}_k\) for each user, SINR  can be expressed by
%\begin{equation}
%\label{eq:9}
%\text{SINR}_k = \frac{|\boldsymbol{\gamma}_k^H \mathbf{x}_k|^2}{\sum_{\substack{m=1 \\ m \neq k}}^{M} |\boldsymbol{\gamma}_k^H \mathbf{x}_m|^2 + \sigma_k^2},
%\end{equation}
%where \(\sigma_k^2\) is the noise power for the \(k\)-th channel. The information rate for User \( k \) can then be calculated as \(R_k = \log_2\left(1 + \text{SINR}_k\right)\).

Given that optimization aim of the downlink beamforming problem is to minimize the total transmit power under the constraints of given sets of SINR for different users, the problem for generic FD case can be expressed as
\begin{subequations}
\label{eq:10}
\begin{align}
&\underset{\mathbf{x}_m, \forall m}{\text{minimize}} \quad \sum_{m=1}^{M} \|\mathbf{x}_m\|^2 \label{eq:10a}\\
& \text{s.t.} \quad \text{SINR}_k = \frac{|\boldsymbol{\gamma}_k^H \mathbf{x}_k|^2}{\sum_{\substack{m=1 \\ m \neq k}}^{M} |\boldsymbol{\gamma}_k^H \mathbf{x}_m|^2 + \sigma_k^2} \geq \delta_k, \quad \forall k. \label{eq:10b}
\end{align}
\end{subequations}
where \(\delta_1, \dots, \delta_k\) represent the SINR thresholds that must be guaranteed for each user, and \(\sigma_k^2\) is the noise power for the \(k\)-th channel. The information rate for User \( k \) can then be calculated as \(R_k = \log_2\left(1 + \text{SINR}_k\right)\).
To achieve a tractable form,  \eqref{eq:10} can be reformulated in the form of convex semidefinite program (SDP) \cite{SDP} as
\begin{subequations}
\label{eq:11}
\begin{align}
 \underset{\mathbf{X}_m}{\text{minimize}} \quad &\sum_{m=1}^{M} \text{Tr}(\mathbf{X}_m)  \label{eq:11a} \\
\text{s.t.}  \quad \text{Tr}(\mathbf{\Gamma}_k \mathbf{X}_k) - \delta_k &\sum_{\substack{m=1 \\ m \neq k}}^{M} \text{Tr}(\mathbf{\Gamma}_k \mathbf{X}_m) - \delta_k \sigma_k^2 \geq 0, \quad \forall k, \label{eq:11b} \\
& \quad \mathbf{X}_m \succeq 0 \, \quad \forall m, \label{eq:11c}
\end{align}
\end{subequations}
where \(\mathbf{X}_m = \mathbf{x}_m \mathbf{x}_m^H
\), \(\mathbf{\Gamma}_k = \boldsymbol{\gamma}_k \boldsymbol{\gamma}_k^H\), and \(\mathbf{X}_m \succeq 0 \) implies that \(\mathbf{X}_m\) is positive semidefinite.
%Problem \eqref{eq:11} is the dual of original problem \eqref{eq:10} as long as the optimal solution  \(\mathbf{X}_m\) is rank-1, i.e., SDP relaxation (SDR) is satisfied. For \eqref{eq:11}, it is known that there exists at least one optimal solution where all \(\mathbf{X}_m\) have rank one. 
The global optimal solutions for original problem \eqref{eq:10} can be obtained by solving \eqref{eq:11} using convex optimization tools such as CVX \cite{cvx1} and performing eigenvalue decomposition (EVD) for \(\mathbf{X}_m\) for all \(m\) (see \cite{SDP} for proof). For FD-based architecture, the problem in \eqref{eq:11} is solved with \( \mathbf{x}_m = \mathbf{w}_m \), and \(\mathbf{X}_m = \mathbf{W}_m\), where \(\mathbf{W}_m = \mathbf{w}_m \mathbf{w}_m^H\). 

\section{PROPOSED BEAMFORMING OPTIMIZATION SOLUTION FOR DMA}
For DMA-based architecture, the optimization problem in \eqref{eq:10} can be revised with \( \mathbf{x}_m = \mathbf{H} \mathbf{Q} \mathbf{w}_m \), which can be derived as
\begin{subequations}
\label{eq:12}
\begin{align}
& \underset{\mathbf{Q}, \mathbf{w}_m, \forall m}{\text{minimize}} \quad \sum_{m=1}^{M} \|\mathbf{H} \mathbf{Q} \mathbf{w}_m\|^2 \label{eq:12a}\\
\text{s.t.} & \quad \frac{|\boldsymbol{\gamma}_k^H \mathbf{H} \mathbf{Q} \mathbf{w}_k|^2}{\sum_{\substack{m=1 \\ m \neq k}}^{M} |\boldsymbol{\gamma}_k^H \mathbf{H} \mathbf{Q} \mathbf{w}_m|^2 + \sigma_k^2} \geq \delta_k, \quad \forall k, \label{eq:12b}\\
& \quad q_{n} \in \mathbb{Q}, \quad \forall n, \label{eq:12c}
\end{align}
\end{subequations}
where \(q_{n}\) is the DMA weight of the \(l\)-th element along the \(i\)-th microstrip, given by \(n=(i-1)N_c + l \), and \(n \in N \).

The optimization problem for the DMA-based architecture, as given in \eqref{eq:12}, differs from the one in \eqref{eq:10}. In the DMA-based optimization problem, the digital precoder vectors \(\mathbf{w}_m\), and the DMA weights \(\mathbf{Q}\) are coupled to each other. Moreover, the amplitudes and phases of the individual elements \( q_{n} \) are dependent on each other, and the phases are limited to the range \([0, \pi]\). These differences make the problem challenging to solve.
The solution of problem \eqref{eq:12} requires joint design of the precoding vector and the DMA weights, where we divide the problem into two decoupled stages to perform optimization for each of the variables individually. Optimization of the digital precoder vectors (\(\mathbf{w}_m\)), and the DMA weights (\(\mathbf{Q}\)) are performed in an alternating manner based on the SDR formulations. 
\subsection{Optimizing the Digital Precoder}
When \(\mathbf{Q}\) is fixed, \eqref{eq:11a} can be rewritten to derive the total transmitted power \(P_{\text{Tx}}\) as follows:
\begin{equation}
\label{eq:13}
\begin{aligned}
P_{\text{Tx}} &= \sum_{m=1}^{M} \text{Tr}(\mathbf{X}_m) = \sum_{m=1}^{M} \text{Tr} \left(\mathbf{H} \mathbf{Q} \mathbf{w}_m (\mathbf{H} \mathbf{Q} \mathbf{w}_m)^H\right) \\
&= \sum_{m=1}^{M} \text{Tr}(\mathbf{Z} \mathbf{W}_m),
\end{aligned}
\end{equation}
where \(\mathbf{Z} = (\mathbf{H} \mathbf{Q})^H \mathbf{H} \mathbf{Q}\).
Moreover, the received power at User \(k\) due to the \(m\)-th beamforming vector of the DMA \((P_{\text{Rx},k,m})\) can be derived as a function of \(\mathbf{W}_m\) according to
\begin{equation}  % Start of subequations environment
\label{eq:14}
\begin{aligned}
P_{\text{Rx},k,m} &= \text{Tr}(\boldsymbol{\gamma}_k^H \mathbf{H} \mathbf{Q} \mathbf{w}_m \mathbf{w}_m^H (\boldsymbol{\gamma}_k^H \mathbf{H} \mathbf{Q})^H) \\ &= \text{Tr}(\mathbf{P}_k \mathbf{W}_m),
\end{aligned}
\end{equation}  % End of subequations environment
where \(\mathbf{P}_k = \boldsymbol{\gamma}_k^H \mathbf{H} \mathbf{Q} (\boldsymbol{\gamma}_k^H \mathbf{H} \mathbf{Q})^H\). Then, combining \eqref{eq:13} and \eqref{eq:14}, optimization problem \eqref{eq:12} can be reformulated in SDP relaxation form as
\begin{equation}
\label{eq:15}
\begin{aligned}
 \underset{\mathbf{W}_m}{\text{minimize}} \quad &\sum_{m=1}^{M} \text{Tr}(\mathbf{Z} \mathbf{W}_m)  \\
\text{s.t.}  \quad \text{Tr}(\mathbf{P}_k \mathbf{W}_k) - \delta_k &\sum_{\substack{m=1 \\ m \neq k}}^{M} \text{Tr}(\mathbf{P}_k \mathbf{W}_m) - \delta_k \sigma_k^2 \geq 0, \quad \forall k, \\
& \quad \mathbf{W}_m \succeq 0, \, \quad \forall m.
\end{aligned}
\end{equation}
After solving the SDP problem defined in \eqref{eq:15}, the digital precoding vectors \(\mathbf{w}_m \in \mathbb{C}^{N_\text{r} \times 1}\) can be obtained via eigenvalue decomposition (EVD) of the associated matrix \(\mathbf{W}_m\) and this solution is global optimum solution if \(\mathbf{W}_m\) is rank-1 (see e.g., \cite{SDP}, Sec. 18.4.2).
\subsection{Optimizing the DMA Weights}
When \(\mathbf{w}_m\) for \(\forall m\) is fixed, the formulations for the optimization of the DMA weights \(\mathbf{Q}\) can be derived by reformulating \eqref{eq:12} in the form of an SDP relaxation, where the variable in the optimization problem is set to be \(\mathbf{Q}\). Utilizing the identity \(\mathbf{A}^T \mathbf{Q} \mathbf{b} = \left(\mathbf{b}^T \otimes \mathbf{A}^T\right) \text{Vec}(\mathbf{Q})\) \cite{DMA_BF1}, \(\mathbf{x}_m\) can be rewritten as:
\begin{equation}
\label{eq:16}
\mathbf{x}_m = \mathbf{H} \mathbf{Q} \mathbf{w}_m = \left(\mathbf{w}_m^T \otimes \mathbf{H}\right) \text{vec}(\mathbf{Q}).
\end{equation}
Then, by defining \(\mathbf{A}_m = \left(\mathbf{w}_m^T \otimes \mathbf{H}\right)^H \in \mathbb{C}^{L \times N}\) and the vector \(\mathbf{q} = \text{vec}(\mathbf{Q}) \in \mathbb{C}^{L \times 1}\), where \(L = N_\text{r}^2 N_\text{c}\), the total transmitted power \(P_{\text{Tx}}\) can be derived as follows:
\begin{equation}
\begin{aligned}
\label{eq:17}
P_{\text{Tx}} = \sum_{m=1}^{M} \text{Tr}\left(\mathbf{X}_m\right) &= \sum_{m=1}^{M} \text{Tr}\left( \mathbf{A}_m^H \mathbf{q} \mathbf{q}^H \mathbf{A}_m \right).
\end{aligned}
\end{equation}
By defining \(\mathbf{\tilde{q}} \in \mathbb{C}^{N \times 1}\), obtained by removing all the zero elements from \(\mathbf{q}\), and \(\mathbf{\tilde{A}}_m \in \mathbb{C}^{N \times N}\), formed by removing the rows corresponding to the indices of the removed elements in \(\mathbf{q}\), \eqref{eq:17} can be rewritten as:
\begin{equation}
\begin{aligned}
\label{eq:18}
P_{\text{Tx}} = \sum_{m=1}^{M} \text{Tr}\left( \mathbf{\tilde{A}}_m^H \mathbf{\tilde{q}} \mathbf{\tilde{q}}^H \mathbf{\tilde{A}}_m \right) &= \sum_{m=1}^{M} \text{Tr}\left( \mathbf{\tilde{B}}_m \mathbf{\tilde{Q}} \right),
\end{aligned}
\end{equation}
where \(\mathbf{\tilde{B}}_m = \mathbf{\tilde{A}}_m \mathbf{\tilde{A}}_m^H\) and \(\mathbf{\tilde{Q}} = \mathbf{\tilde{q}} \mathbf{\tilde{q}}^H\).

Following the derivations for \(P_{\text{Tx}}\), 
the received power at User \(k\) due to the \(m\)-th beamforming vector of the DMA \(P_{\text{Rx},k,m}\) can be derived similarly.  Given the fact that \(\mathbf{a}^T \mathbf{Q} \mathbf{b} = \left(\mathbf{b}^T \otimes \mathbf{a}^T\right) \text{vec}(\mathbf{Q})\), \( y_{k,m} \) given in \eqref{eq:8} can be defined and reformulated as
\begin{equation}
\begin{aligned}
\label{eq:19}
y_{k,m} &= \left(\mathbf{w}_m^T \otimes (\boldsymbol{\gamma}_k^H \mathbf{H})\right) \text{vec}(\mathbf{Q})= \mathbf{c}_{k,m}^H \mathbf{q},
\end{aligned}
\end{equation}
where \(\mathbf{c}_{k,m} = \left(\mathbf{w}_m^T \otimes (\boldsymbol{\gamma}_k^H \mathbf{H})\right)^H \in \mathbb{C}^{L \times 1}\). Furthermore, the modified vectors, obtained by removing zero elements as described above, can be defined as \(\mathbf{\tilde{c}}_{k,m} \in \mathbb{C}^{N \times 1}\) and \(\mathbf{\tilde{q}} \in \mathbb{C}^{N \times 1}\). Then,  \(P_{\text{Rx},k,m}\) can be further simplified as:
\begin{equation}
\label{eq:20}
\begin{aligned}
P_{\text{Rx},k,m} &= \text{Tr}\left(\mathbf{\tilde{c}}_{k,m}^H \mathbf{\tilde{q}}(\mathbf{\tilde{c}}_{k,m}^H \mathbf{\tilde{q}})^H\right)= \text{Tr}\left(\mathbf{\tilde{C}}_{k,m} \mathbf{\tilde{Q}}\right),
\end{aligned}
\end{equation}
\vspace{0.01in} where \(\mathbf{\tilde{C}}_{k,m} = \mathbf{\tilde{c}}_{k,m} \mathbf{\tilde{c}}_{k,m}^H\).

Finally, combining \eqref{eq:18} and  \eqref{eq:20} the optimization problem is formulated as 
\begin{equation}
\label{eq:21}
\begin{aligned}
 \underset{\mathbf{\tilde{Q}}}{\text{minimize}} \quad &\sum_{m=1}^{M} \text{Tr}(\mathbf{\tilde{B}}_m \mathbf{\tilde{Q}}) \\
\text{s.t.}  \quad \text{Tr}(\mathbf{\tilde{C}}_{k,k} \mathbf{\tilde{Q}}) - \delta_k &\sum_{\substack{m=1 \\ m \neq k}}^{M} \text{Tr}(\mathbf{\tilde{C}}_{k,m} \mathbf{\tilde{Q}}) - \delta_k \sigma_k^2 \geq 0, \quad \forall k, \\
& \quad \mathbf{\tilde{Q}} \succeq 0.
\end{aligned}
\end{equation}

For given digital precoding vectors \(\{\mathbf{w}_m^\star\} \, \forall m\), the solution of \eqref{eq:21} yields the optimal matrix \(\mathbf{\tilde{Q}}^\star \in \mathbb{C}^{N \times N}\). The vector \(\mathbf{\tilde{q}}^\star \in \mathbb{C}^{N \times 1}\) can then be obtained via the EVD of \(\mathbf{\tilde{Q}}^\star\). The special case solution of problem \eqref{eq:12} can be defined when the condition in \eqref{eq:12c}, i.e, \(q_{n} \in \mathbb{Q}\) is relaxed to \(q_{n} \in \mathbb{C}^{N \times 1}\). We refer to this scenario as "unrestricted weights" scenario, since the amplitudes and phases of the weights can be optimized freely within the complex domain (\(\mathbb{C}^{N \times 1}\)). However, when performing optimization with a DMA-based architecture using \eqref{eq:12}, the solution vector \(\mathbf{\tilde{q}}^\star\) must be mapped onto the Lorentzian circle described in \eqref{eq:7p}. For this mapping, we use the method described in \cite{DMA1}, specifically, the entry-wise projection that aims to find the nearest points on the Lorentzian circle, \(\tilde{q}_{n} \in \mathbf{\tilde{q}}\), from the elements of the solution vector \(\tilde{q}^\star_{n} \in \mathbf{\tilde{q}}^\star\). The problem \eqref{eq:21} can be considered as a relaxation of the problem in \eqref{eq:12} based on the Lorentzian constraints without mapping. The Lorentzian Mapping step ensures that the optimized DMA weights comply with the constraints in \eqref{eq:7p}, where their amplitudes and phases follow the Lorentzian relation, limiting the optimization space and beamforming performance.

%According to this relation, the amplitude of the DMA weights for individual elements is determined by the corresponding phase shift of those elements, which is restricted to the range \([0, \pi]\). As a result, these constraints lead to a loss of performance.

Finally, solving the individual problems \eqref{eq:15} and \eqref{eq:21} in an alternating way effectively leads to the convergence of the digital precoder vectors toward the optimal solution, while optimizing the DMA weights toward suboptimal solutions. This approach is encapsulated in the proposed algorithm, which is detailed in Algorithm \ref{alg:proposed_algorithm}. The worst-case complexity of the proposed algorithm can be determined in reference to the complexity of SDP using an interior point method, which is given with \( O(\max\{m, n\}^{4} n^{\frac{1}{2}} \log(1/\epsilon)) \), where \( m\) is the number of constraints, \( n \) denotes the number of optimization variables, and \( \epsilon \) represents the accuracy \cite{5447068}. In this, the number of optimization variables can be determined with \( n = N_\text{r}\) for SDP problem in \eqref{eq:15} and \( n = N\) for SDP problem in \eqref{eq:21}, where the number of constraints \( m\) in both SDP problems  is equal to the number of users (\( K\)).  Therefore, the overall complexity remains polynomial in both the problem size \( n \), and the number of constraints \( m\), ensuring computational feasibility.

\begin{algorithm}
\caption{Proposed Algorithm for Solving Problem \eqref{eq:12}}
\label{alg:proposed_algorithm}
\begin{algorithmic}[b]
\State \textbf{Initialize:} $\mathbf{Q}^{(0)}$;
    \State Solve \eqref{eq:15} to compute $\left\{\mathbf{w}^{(0)}\right\}_{m=1}^{M}$ and $P_{\text{Tx}}^{(0)}$;
%\State Update $\left\{\mathbf{w}^{(0)}\right\}_{m=1}^{M}$ and $P_{\text{Tx}}^{(0)}$;
\For{$t = 1, \dots, T$}
    \State Solve \eqref{eq:21} to obtain $\mathbf{\tilde{Q}}^\star$ and compute $\mathbf{\tilde{q}}^\star$;
    \State Apply Lorentzian Mapping to obtain $\mathbf{\tilde{q}}$;
    \State Update $\mathbf{Q}^{(t)}$ using $\mathbf{\tilde{q}}$ and \eqref{eq:7};
    \State Solve \eqref{eq:15} to compute $\left\{\mathbf{w}^{(t)}\right\}_{m=1}^{M}$ and $P_{\text{Tx}}^{(t)}$;

    \If {$P_{\text{Tx}}^{(t)} \leq P_{\text{Tx}}^{(t-1)}$}
        \State Update $\left\{\mathbf{w}^{(f)}\right\}_{m=1}^{M}$, $\mathbf{Q}^{(f)}$, and $P_{\text{Tx}}^{(f)}$;
    \EndIf
\EndFor
\State \textbf{Output:} $\left\{\mathbf{w}^{(f)}\right\}_{m=1}^{M}$, $\mathbf{Q}^{(f)}$, $P_{\text{Tx}}^{(f)}$.
\end{algorithmic}
\end{algorithm}

\subsection{Discussion}
When comparing the optimization problems for FD-based architecture \eqref{eq:10} and DMA-based architecture \eqref{eq:12}, DMA has limitations in terms of reduced DoF due to the Lorentzian constraint in \eqref{eq:12c} and the reduced number of optimization variables. The fewer optimization variables result from the reduced number of RF chains in DMA compared to FD, which helps lower transmitter power consumption. However, the reduced DoF may degrade beamforming performance and increase the required transmission power. This study focuses on quantifying the performance gap caused by these DoF limitations in DMA. We assess their impact on beamforming through numerical comparisons of optimization problems. The power consumption of the RF chain will be examined in future work. 
For comparison purposes, we consider two optimization problems OP1 and OP2. When the problem given in \eqref{eq:11} is solved with the same antenna spacing of DMA configuration, the optimization problem is called OP1. 
Note that, OP1 corresponds to the solution of the FD-based architecture when $d_x = \lambda/2$. %OP1 represents a physically non-realizable boundary for DMA optimizations, serving as a benchmark with unconstrained DMA weights and the maximum number of degrees of freedom (DoF) in terms of optimization parameters.
The number of optimization variables in OP1 is given by $\text{nDoF} = N_\text{r}N_\text{c}M$, whereas for DMA, it is $N_\text{r}(M + N_\text{c})$ (from \eqref{eq:15} and \eqref{eq:21}). This suggests that as the number of users increase, DMA will leverage less DoF with respect to OP1, leading to a performance gap in terms of beamforming performance. %Hence, comparisons of DMA optimization with OP1 and FD will highlight these impacts. 
Furthermore, we introduce OP2, where the DMA optimization problem is solved with unrestricted weights to further analyze the impact of Lorentzian constraints on DoF reduction in DMA. Numerical comparisons with OP1 and OP2 will illustrate the impact of DoF limitations and Lorentzian constraints on beamforming performance.

\section{NUMERICAL RESULTS}

In the numerical results, we consider the simulation setup that a uniform planar array is placed in the \textit{xy}-plane, and the users are randomly distributed over 200\footnote{The near-field channel model used in this paper and in the related literature is deterministic, excluding randomness like  Rayleigh fading. Prior studies, e.g., \cite{DMA_BF1, Azarbahram_2024}, evaluate DMA performance for limited user positions with up to 30 realizations. Here, Monte Carlo simulations are increased to 200 to assess the transmitter performance across a region with randomly located users.} realizations within the half-circles with radius $r$ lie in the \textit{xz}-plane, covering both near-zone (\text{$0.1d_\text{F} < r < 1d_\text{F}$}) and far-zone regions (\text{$1d_\text{F} < r < 5d_\text{F}$}). Here, $d_\text{F}$ is the Fraunhofer distance, i.e., \(d_\text{F} \triangleq \frac{2D^2}{\lambda}\) for antenna aperture length of \(D\) and wavelength \({\lambda}\). Throughout the experimental study, the frequency is set to \( f = 28 \, \text{GHz} \), and the aperture size is consistently fixed at \( 10 \, \text{cm} \times 10 \, \text{cm} \) for all structures. Optimizations are performed to guarantee the minimum achievable rate for each user (\( R_{\text{min},k} \)), which can be related to the SINR conditions in optimization problems by \(\delta_k \triangleq (2^{R_{\text{min},k}} - 1)\),  where \( R_{\text{min},k}\) is the same for all \(k\). The noise power is set to \( \sigma_k^2 = -114 \, \text{dBm} \).

For DMA, flat frequency characteristics are achieved by assuming the \( \mathbf{H} \) matrix is an identity matrix. For all structures, the separation between rows of the UPA is kept constant at \(d_y = \lambda/2\), where \( N_\text{r} = \left\lfloor 2D/\lambda \right\rfloor\). The separation between columns of the UPA, denoted by \( d_x \), is varied for the case of DMA, while it is fixed at \(d_x = \lambda/2\) for the FD-based architecture as a benchmark.

\begin{figure}[!t]
\centering
\includegraphics[width=0.35\textwidth]{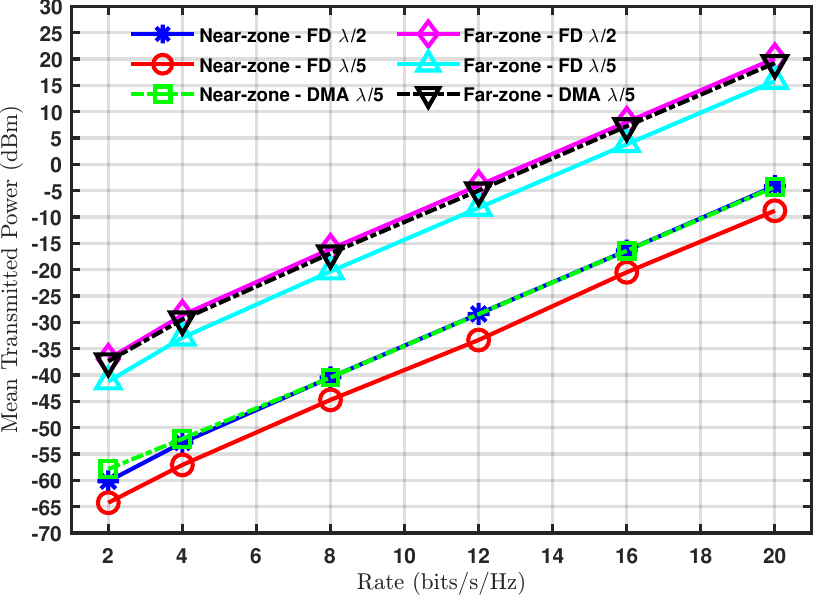}
\caption{Mean transmitted power versus minimum rate requirement at $K=2$.}
\label{fig:results_1}
\end{figure}

%To demonstrate the robustness of the proposed optimization algorithm for DMA, we perform a performance analysis of the mean transmit power as a function of the minimum rate constraint \( R_{\text{min},k} \) for \( K=2 \) user realizations in different zones (Zone 1: near-zone, Zone 2: far-zone), as given in Fig. \ref{fig:results_1}. Three optimization scenarios are considered: (1) DMA with \(d_x = \lambda/5\), (2) FD with \(d_x = \lambda/2\), and (3) OP1 with \(d_x = \lambda/5\), highlighting the optimal boundary for DMA in terms of DoF. As expected, the mean transmitted power increases linearly as the minimum rate requirement increases. While DMA with \(d_x = \lambda/5\) and FD with \(d_x = \lambda/2\) performs very similar, DMA requires, on average, 4.2 dBm and 3.4 dBm higher transmit power compared to OP1 with \(d_x = \lambda/5\) in the near-zone and far-zone, respectively. The performance gap of DMA compared to optimal boundary OP1 remains constant in both regions with increasing SINR, which is the efect of the reduced number of DoF for DMA.
To evaluate the robustness of the proposed DMA optimization, we analyze mean transmit power as a function of the minimum rate constraint \( R_{\text{min},k} \) for \( K=2 \) users in near-zone  and far-zone, as shown in Fig. \ref{fig:results_1}. Three scenarios are considered: (1) DMA with \(d_x = \lambda/5\), (2) FD with \(d_x = \lambda/2\), and (3) OP1 with \(d_x = \lambda/5\), representing the optimal boundary for DMA in terms of DoF. As expected, mean transmit power increases linearly with the minimum rate requirement. While DMA (\(d_x = \lambda/5\)) and FD (\(d_x = \lambda/2\)) perform similarly, DMA requires 4.2 dBm and 3.4 dBm higher transmit power than OP1 in the near-zone and far-zone, respectively. The performance gap between DMA and OP1 remains constant across both regions with increasing SINR, demonstrating the impact of reduced DoF in DMA.

\begin{figure}[!t]
\centering
\includegraphics[width=0.35\textwidth]{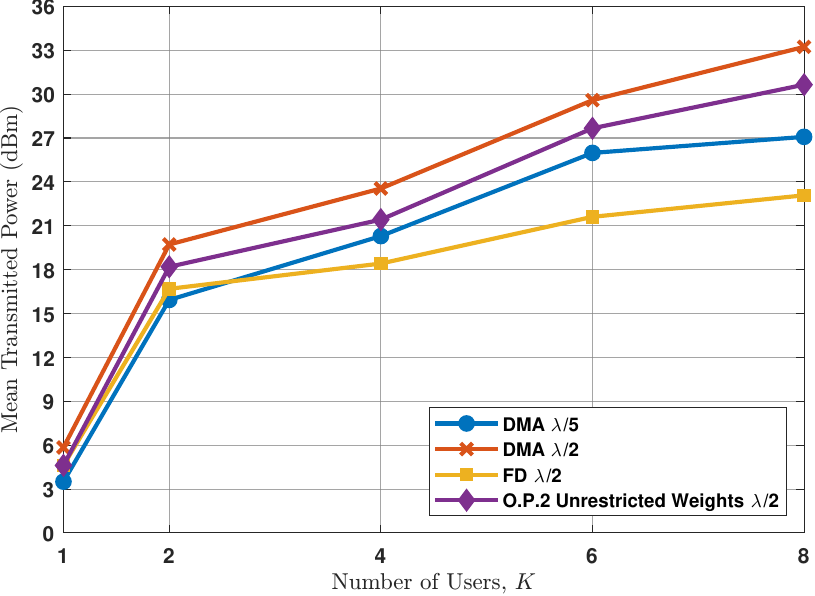}
\caption{Mean transmitted power versus number of users ($K$) at $R_{\text{min},k}=$20 bps/Hz for $\forall k$.}
\label{fig:results_2}
\end{figure}

Next in Fig. \ref{fig:results_2}, we examine the mean transmit power versus the number of users, with a minimum rate constraint \( R_{\text{min},k} = 20 \) bits/s/Hz\footnote{The selected rate parameter is in align with prior DMA-aided near-field LoS studies \cite{DMA_BF1}, while the performance gap interpretation remains valid also for lower rate values, shown in Fig. \ref{fig:results_1}.}  for all users. In this analysis, 200 realizations are performed for a region covering both the near-zone and far-zone users. Optimization scenarios, i.e., DMA, FD, and OP2- Unrestricted Weights (UW) are considered, all with the same antenna spacing \(d_x = \lambda/2\) to ensure the same number of antennas across all structures. Additionally, DMA with \(d_x = \lambda/5\) is included for comparison purposes. It is observed that the performance gap between DMA and FD with the same antenna spacing \(d_x = \lambda/2\), gradually increases as number of users increases. This can be related to the effect of DoF, as discussed in  Section III.C. Numerically, the difference in mean transmit power demands grows from 2.3 dB for \( K=1 \) to 10.2 dB for \( K=8 \). Meanwhile, DMA with \(d_x = \lambda/5\), offers better performance in terms of mean transmit power for \(K=1,2\) than FD, while FD with \(d_x = \lambda/2\) outperforms it for a higher number of users, i.e., (\(K>2\)). This yields that DMA requires more elements and DoF than it is with \(d_x = \lambda/5\) to achieve the same performance as FD under these conditions for \(K>2\). 

\begin{figure}[t]
    \centering
    \begin{subfigure}[b]{0.35\textwidth}
        \centering
        \includegraphics[width=\textwidth]{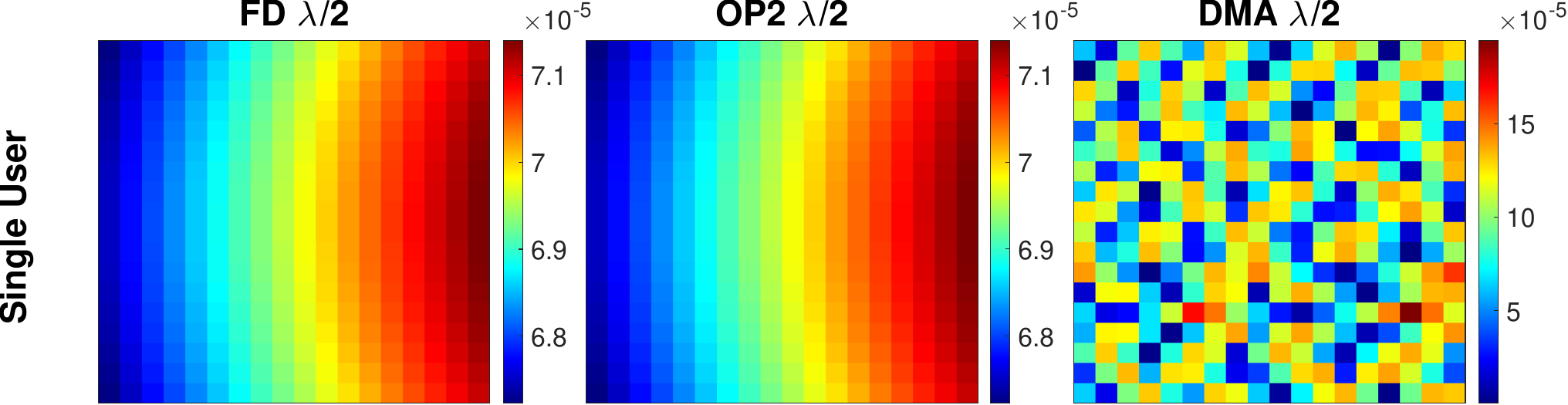}
        \caption{}
        \label{fig:results_4a}
        \vspace{0.1cm}
    \end{subfigure}
    \begin{subfigure}[b]{0.35\textwidth}
        \centering
         \includegraphics[width=\textwidth]{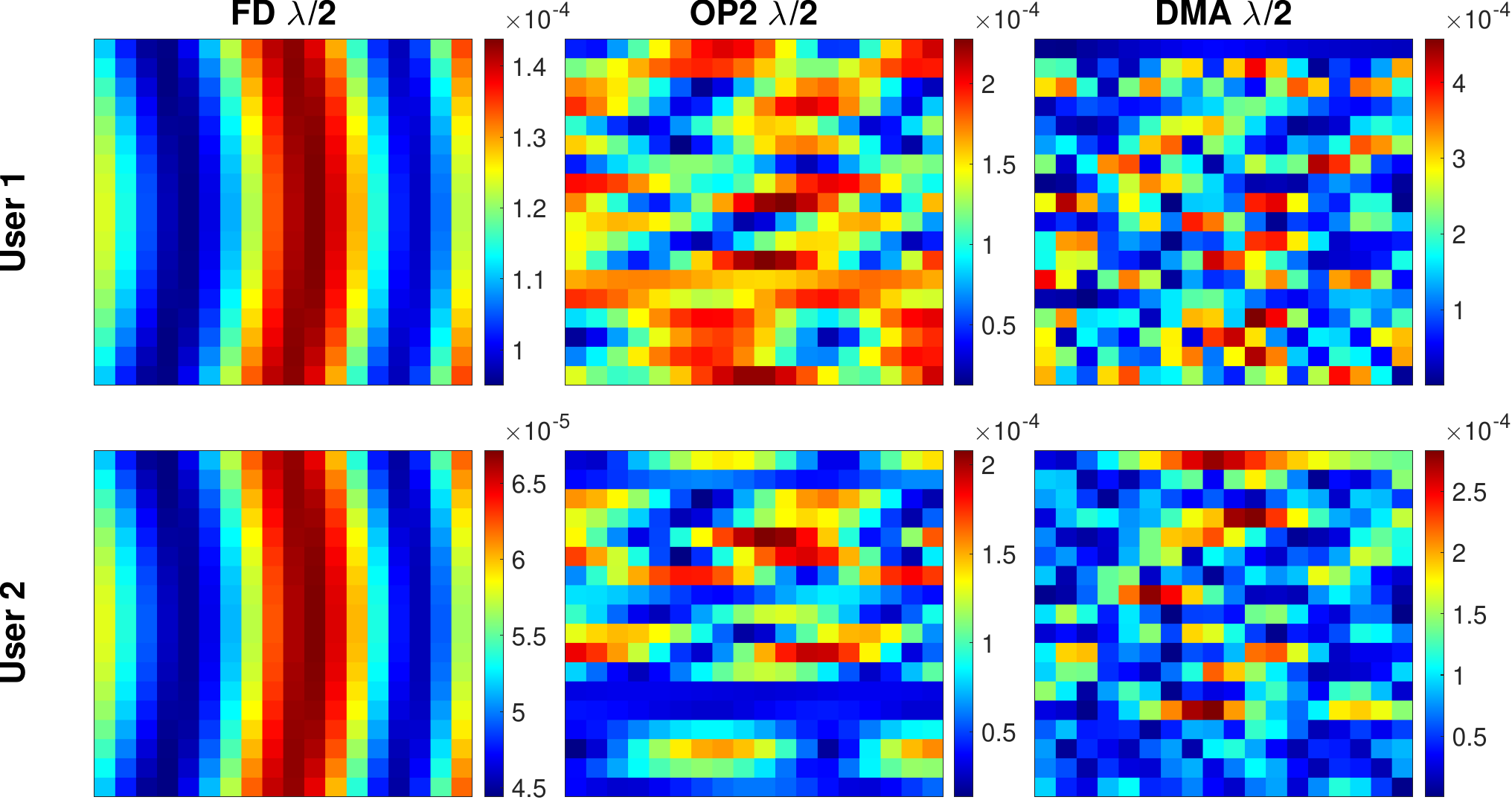}
        \caption{}
        \label{fig:results_4b}
    \end{subfigure}
    \caption{Comparison of different aperture weight distributions for optimizations OP1-FD, OP2-UW, and OP2-DMA   $d_x=d_y=\lambda/2$; (a) ($K=1$), (b) ($K=2$).}
    \label{fig:results_4}
\end{figure}

To examine the effect of DoF for both Lorentzian mapping and number of optimization variables, we can further compare the results of DMA and FD with OP2-UW under the same antenna spacing at \(d_x = \lambda/2\). The comparison between OP2-UW and DMA highlights the impact of restrictions on DMA weights with Lorentzian mapping. Specifically, the performance gap is approximately 1.1 dBm for \( K=1 \), increasing to 2.6 dBm as \( K \) reaches to 8. On the other hand, comparison of FD and OP2-UW can be interpreted in terms of impact of number of optimization variables on achievable transmit power. For \( K=1\), FD and OP2-UW achieve nearly the same mean transmit power (4.62 dBm), indicating that the proposed optimization algorithm for DMA performs robustly. Algorithm \ref{alg:proposed_algorithm} for solving \eqref{eq:12} yields the same result as the optimal solution of \eqref{eq:11} when the DoF for both problems are the same. However, as \( K \) increases to 2 and beyond, FD begins to outperform OP2-UW, benefiting from a larger number of optimization variables. To further support this, Fig. \ref{fig:results_4} presents the aperture weight distributions for three scenarios: DMA, FD, and OP2-UW, each with \(d_x = \lambda/2\). For this analysis, a single realization among the 200 simulations for \( K=1 \) and \( K=2 \) was selected.  For \( K=1 \), both FD and OP2-UW provide the same aperture weight distributions for this particular case. However, the distribution differs for the DMA case due to the constraints imposed by the Lorentzian restrictions. On the other hand, For \( K=2\), even OP2-UW cannot provide aperture weight distribution obtained with FD, which shows the effect of lower DoFs.

\begin{figure}[!t]
\centering
\includegraphics[width=0.35\textwidth]{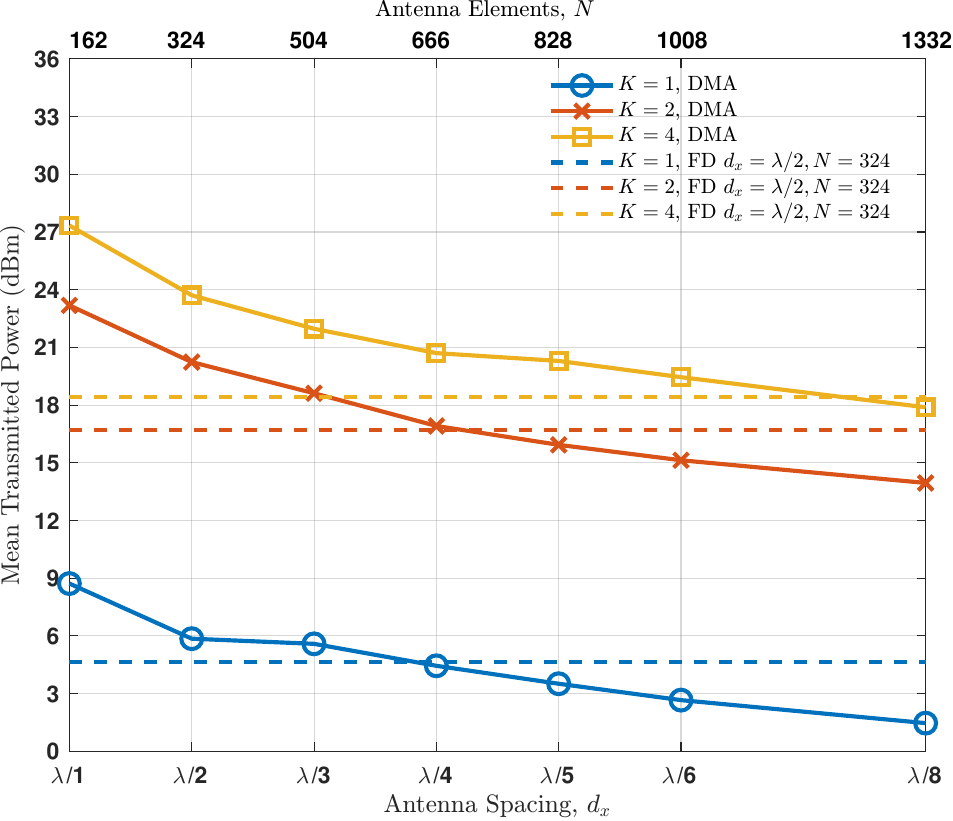}
\caption{Mean transmitted power versus antenna spacing ($d_x$), and number of elements ($N$) at $K = 1$, $K = 2$, and $K = 4$.}
\label{fig:results_5}
\end{figure}

Finally, in Fig. \ref{fig:results_5}, we present the results analyzing the effect of number of elements (\(N\)) in DMA to achieve the same performance as conventional FD-based architectures with \(d_x = \lambda/2\), yielding \(N=324\). Simulations were conducted with a minimum rate constraint \( R_{\text{min},k} = 20 \) bits/s/Hz, separately for different numbers of users \( K=1 \), \( K=2 \), and \( K=4 \), and for various antenna spacings \( d_x \) of the DMA. For a fair comparison, the aperture size was kept constant, where smaller values of \( d_x \) resulted in a higher number of antennas per row \( N_\text{c} = \left\lfloor {D}/{d_x} \right\rfloor \). As expected, the results in Fig. \ref{fig:results_5} demonstrate that as the number of elements (\(N\)) increases, the performance of the DMA gradually improves, and the performance gap relative to FD narrows. Numerically, for \( K=1 \), DMA surpasses FD performance at \(N=666\), and for \( K=2 \), at nearly \(N=666\). For \( K=4 \), the performance gap relative to FD decreases as \(N\) increases, from 8.9 dB at \(N=162\) to 2.3 dB at \(N=666\), with DMA outperforming FD beyond \(N=1008\) . It is important to note that DMA offers additional efficiency gains due to lower power consumption from its reduced number of RF chains, although this aspect is not considered in this analysis. Specifically, DMA requires significantly fewer RF chains, with \( N_\text{r} = 18 \) for \(d_y = \lambda/2\) , compared to FD with \( N = N_\text{r} \times N_\text{c} = 818 \) for \(d_x = \lambda/2\) and \(d_y = \lambda/2\). These results suggest that when using DMA for beamforming applications, as the number of users increase, \(N\) should be also increased to alleviate the impact of reduced DoF, while using the same number of RF chains (\(N_\text{r}\)).

\section{Conclusion}

In this work, we investigated DMA-aided multi-user MISO systems under minimum achievable rate constraints with the joint optimization of DMA weights and digital precoders, aiming to minimize transmitted power. We demonstrated that the joint optimization for DMA yields the same result with FD when the DoF for both scenarios are the same, i.e., specifically in the case of $K=1$ and without any Lorentzian constraints on DMA weights. For higher number of users and simulation with Lorentzian constraints of DMA weights, the performance gap is presented due to the reduced DoF for DMA. Moreover, impact of Lorentzian constraints on this gap is shown to slightly increase as the number of users within the network increases, whereas this gap remains constant for higher requirement of achievable rate in the same setup. On the other hand, impact of the reduction in the number of optimization variables has significant impact, leading to higher increase in the performance gap with the increase of number of users. It has also been shown that, as the number of users increases, higher antenna array elements are required for DMA to outperform FD with the same number of RF chains. Considering these insights, the power consumption of RF chains, mutual coupling effects, and the exploration of DoF for systems with a massive number of devices in DMA-aided configurations could be potential topics for future research.

\section*{Acknowledgment}
This work was supported by UK Research and Innovation under the UK government’s
Horizon Europe funding guarantee through MSCA-DN SCION Project Grant
Agreement No.101072375 [Grant Number: EP/X027201/1].

\bibliography{main}



%\bibliographystyle{IEEEtran}

\end{document}